\documentclass[conference]{IEEEtran}
\IEEEoverridecommandlockouts

\usepackage{cite}
\usepackage{amsmath,amssymb,amsfonts}
\usepackage{algorithmic}

\usepackage{graphicx}
\usepackage{textcomp}
\usepackage{multirow}   
\usepackage{subcaption}
\usepackage[colorlinks=true,
            linkcolor=black,   
            citecolor=blue,   
            urlcolor=blue]     
            {hyperref}
\usepackage{array}
\usepackage{dsfont}
\usepackage{rotating}
\usepackage{dsfont}
\usepackage{diagbox}
\makeatletter
\newcommand\blfootnote[1]{%
  \begingroup
  \renewcommand\thefootnote{}
  \footnotetext{#1}%
  \addtocounter{footnote}{-1}
  \endgroup
}
\makeatother
\usepackage{siunitx} 
\usepackage{url}
\usepackage{xcolor}
\DeclareMathOperator{\vect}{vec}
\DeclareMathOperator*{\argmin}{\arg\!\min}
\def\BibTeX{{\rm B\kern-.05em{\sc i\kern-.025em b}\kern-.08em
    T\kern-.1667em\lower.7ex\hbox{E}\kern-.125emX}}
\begin{document}

\title{Fast Single-Snapshot Harmonic Recovery with 2D Sparse Arrays using BCCB Matrices \\
}

\author{\IEEEauthorblockN{Youval Klioui}
\IEEEauthorblockA{\textit{Dept. of Electrical Engineering} \\
\textit{Eindhoven University of Technology}\\
Eindhoven, The Netherlands \\                                                   
y.klioui@tue.nl}

}

\maketitle

\begin{abstract}
	We introduce an efficient implementation of sparse recovery methods for the problem of harmonic estimation with 2D sparse arrays using a single snapshot. By imposing a uniformity constraint on the harmonic grids of the subdictionaries used in the sparse recovery problem, in addition to a mild constraint on the array topology that consists in having the elements lie on a grid specified in half-wavelength units, we show that the Gram matrices that appear in these sparse recovery methods exhibit a block-circulant with circulant blocks (BCCB) structure. The BCCB structure is then exploited to reduce the computational complexity of the matrix-vector products that appear in these methods through the use of 2D fast Fourier transforms (FFT) from $\mathcal{O}((L_{1}L_{2})^{2})$ down to $\mathcal{O}(L_{1}L_{2}\log(L_{1}L_{2}))$ operations per iterations, where $L_{1}, L_{2}$ are the lengths of the subdictionaries used for estimating the harmonics in the first and second dimension, respectively. We experimentally verify the proposed implementation using the iterative shrinkage thresholding algorithm (ISTA), the fast iterative shrinkage-thresholding algorithm (FISTA), and the alternating direction method of multipliers (ADMM) where we observe improvements in the runtime of up to two orders of magnitude.
\end{abstract}

\begin{IEEEkeywords}
Harmonic Recovery, 2D Sparse Arrays, Single-Snapshot, Compressed Sensing, BCCB Matrices 
\end{IEEEkeywords}

\section{Introduction}
\label{intoduction}
Efficient 2D harmonic estimation with sparse arrays using a single snapshot remains a challenging problem. The sparse nature of the array renders the use of beamformers challenging as they suffer from high sidelobes in addition to having an inherently limited resolution set by the Rayleigh limit. Subpspace-based methods such as estimation of signal parameters through rotation invariance (ESPRIT) \cite{esprit}\cite{2desprit1}\cite{2desprit2} or multiple signal classification (MUSIC) \cite{music}\cite{2dmusic1}\cite{2dmusic2} require spatial smoothing\cite{smoothing} under the single-snapshot scenario, which in turn requires the subarrays used in the spatial smoothing algorithm to have an identical topology, a requirement that would be difficult to meet without additional constraints on the positioning of the elements of the 2D array. Maximum likelihood estimation \cite{mle} also proves prohibitively expensive under the 2D setting as an exhaustive search over a multidimensional grid that grows with the number of sources is required. Sparse recovery methods such as the least absolute shrinkage and selection operator (LASSO)\cite{lasso}, although able to provide high resolution using a single snapshot and without imposing stringent requirements on the array geometry, rely on solvers such as the iterative-shrinkage thersholding (ISTA)\cite{ista}, the fast iterative shrinkage-thresholding algorithm (FISTA)\cite{fista}, or the alternating direction method of multipliers (ADMM)\cite{admm} that are computationally intensive in the 2D setting as they require $\mathcal{O}((L_{1}L_{2})^{2})$ operations per each iteration, where $L_{1}$ and $L_{2}$ are the lengths of the dictionaries used for estimating the harmonics in the first and second dimension, respectively.  By exploiting the block-circulant with circulant blocks (BCCB) matrices that arise in these iterative methods, it can be shown that the computational cost can be reduced to $\mathcal{O}(L_{1}L_{2}\log(L_{1}L_{2}))$.

\blfootnote{The repository for replicating the results reported here can be found at:\\
\url{https://github.com/youvalklioui/sparse-2d-array-bccb}}

\section{Problem Formulation}
\label{problem formulation}
\subsection{2D Sparse Array Signal Model}
We first consider a planar uniform array (URA) with elements located on a 2D grid specified by the coordinates $(x_{m_{1}}, y_{m_{2}},0)$ and expressed in units with respect to the wavelength $\lambda$, i.e. $(x_{m_{1}},y_{m_{2}})=(m_{1}\gamma_{1}\lambda,m_{2}\gamma_{2}\lambda)$ for $m_{i} = 0, 1, \hdots, M_{i}-1$, and $\gamma_{i} >0$, with $i=1,2$. Such an array will have an aperture of $(M_{1}-1)\gamma_{1}\lambda$ and $(M_{2}-1)\gamma_{2}\lambda$ in the $x$ and $y$ directions, respectively, and will contain $M_{1}M_{2}$ elements. The resulting single-snapshot  signal model for $K$ targets impinging on the URA can be formulated as
\begin{align}
	\mathbf{Y}=\sum_{k=1}^{K}c_{k}\mathbf{A}(\phi_{k},\theta_{k})+\mathbf{N}
	\label{eq:signal_model}
\end{align}
where $\mathbf{Y} \in \mathbb{C}^{M_{1} \times M_{2}}$ is the measurement single-snapshot signal received on the URA, $\mathbf{N}(m_{1},m_{2}) \sim \mathcal{CN}(0, {\sigma}^{2})$ represents normally distributed additive noise, and  $\mathbf{A}(\phi_{k},\theta_{k})\in \mathbb{C}^{M_{1} \times M_{2}}$ represents the steering matrix corresponding to the $k$-th target with amplitude $c_{k}\in \mathbb{C}$ located at angular coordinates $(\phi_{k}, \theta_{k})$, for $\phi_{k} \in [-\pi, \pi[$ and $\theta_{k} \in [0, \frac{\pi}{2}[$, and is explicitly given by \cite{trees2002optimum}
\begin{align}
	\mathbf{A}(\phi_{k},\theta_{k})(m_{1},m_{2}) &= \exp\big(-j2\pi \big(m_{1}\gamma_{1}\cos(\phi_{k})\sin(\theta_{k})+ \nonumber \\ &m_{2}\gamma_{2}\sin(\phi_{k})\sin(\theta_{k})\big)\big) \nonumber \\
	&= \exp\big(-j2\pi \big(m_{1}f_{1,k}^{*}+ m_{2}f_{2,k}^{*}\big)\big),
\end{align}
where we have defined the tuple of harmonics $(f_{1,k}^{*}, f_{2,k}^{*}) \triangleq (\gamma_{1}\cos(\phi_{k})\sin(\theta_{k}), \gamma_{2}\sin(\phi_{k})\sin(\theta_{k}))$. From a set of estimates for the harmonics $\{(f_{1,k}^{*}, f_{2,k}^{*})\}_{k=1}^{K}$ we can directly obtain a corresponding set of estimates for the angle coordinates $\{(\phi_{k}, \theta_{k})\}_{k=1}^{K}$. With the ranges specified for $\theta_{k}$ and $\phi_{k}$, we have $-\gamma_{i}\le f_{i,k}^{*}<\gamma_{i}$, for $i=1,2$. We will next layout the sparse recovery framework for obtaining estimates of $\{(f_{1,k}^{*}, f_{2,k}^{*})\}_{k=1}^{K}$. By applying the vectorization operator on both sides of \eqref{eq:signal_model}, we can rewrite the signal model as
\begin{align}
	\mathbf{y}=\sum_{k=1}^{K}c_{k}\mathbf{a}_{2}(f_{2,k}^{*})\otimes \mathbf{a}_{1}(f_{1,k}^{*})+\mathbf{n},
	\label{eq:vec_signal_model}
\end{align}
where $\mathbf{y} = \textrm{vec}(\mathbf{Y})$, $\mathbf{n}=\textrm{vec}(\mathbf{N})$, and $\mathbf{a}_{i}(f_{i,k}^{*})(m_{i})=\exp(-j2\pi f_{i,k}^{*} m_{i})$, for $i=1,2$, and the resulting dimensions are given by $\mathbf{y}, \mathbf{n}  \in \mathbb{C}^{M_{2}M_{1}}$, $\mathbf{a}_{i}(f_{i,k}^{*})\in \mathbb{C}^{M_{ i}}$, for $i = 1, 2$. Under the sparse recovery setting such as the LASSO, the measurement vector $\mathbf{y}$ is expressed using the following forward model \cite{lasso}\cite{candesrobust}
\begin{align}
	\mathbf{y}=\mathbf{D}\mathbf{c}+\mathbf{n},
	\label{eq:forward_model}
\end{align}
where $\mathbf{c} \in \mathbb{C}^{L_{2}L_{1}}$ is a sparse vector we would like to estimate since its support corresponds to estimates of the harmonics $\{(f_{1,k}^{*}, f_{2,k})\}_{k=1}^{K}$, with $K<<L_{1}L_{2}$, and $\mathbf{D} \in \mathbb{C}^{M_{1}M_{2}\times L_{1}L_{2}}$ is the so-called dictionary matrix which, following \eqref{eq:vec_signal_model}, is explicitly given by $\mathbf{D} = \mathbf{D}_{2}\otimes\mathbf{D}_{1}$, where in turn $\mathbf{D}_{i} \in \mathbb{C}^{M_{i}\times L_{i}}$ is a subdictionary of length $L_{i}$ used for estimating the $K$ harmonics in the $i$-th dimension, $\{f_{i,k}^{*}\}_{k=1}^{K}$, and is explicitly given by
\begin{align}
\mathbf{D}_{i} (m_{i}, l_{i}) = \exp(-j2\pi f_{i,l_{i}}m_{i}),
\label{eq:subdictionary}
\end{align} 
for $m_{i} = 0, 1, \hdots, M_{i}-1$, $l_{i} = 0, 1, \hdots L_{i}-1$, and $i=1,2$. We should note that the $l_{i}$-th column of $\mathbf{D}_{i}$ is simply $\mathbf{a}_{i}$ defined in the signal model in \eqref{eq:vec_signal_model} evaluated at $f_{l_{i}}$. Thus $\mathbf{D}_{1}$ and $\mathbf{D}_{2}$ are the direct result of evaluating $\mathbf{a}_{1}(f_{1})$ and $\mathbf{a}_{2}(f_{2})$ on the harmonic grids given by $\mathcal{F}_{1} = \{f_{1,0}, f_{1,1}, \hdots, f_{1,L_{1}-1}\}$ and $\mathcal{F}_{2}=\{f_{2,0}, f_{2,1}, \hdots, f_{2,L_{2}-1}\}$, respectively. As $\mathbf{D}$ is obtained by taking the Kronecker product of $\mathbf{D}_{2}$ and $\mathbf{D}_{1}$, we can see that each column of $\mathbf{D}$ therefore corresponds to a specific tuple of harmonics $(f_{1,l_{1}},f_{2,l_{2}})$ sampled from $\mathcal{F}_{1}$ and $\mathcal{F}_{2}$, and there are exactly $L_{1}L_{2}$ such possible combinations. To be more specific, the tuple $(f_{1,l_{1}},f_{2,l_{2}})$ corresponds to the $(l_{1}+l_{2}L_{1})$-th column of $\mathbf{D}$, and this column is exactly equal to $\mathbf{a}_{2}(f_{l_{2}})\otimes \mathbf{a}_{1}(f_{l_{1}})$. Additonally, $\mathbf{c}(l_{1}+l_{2}L_{2})$ corresponds to the amplitude of the harmonics tuple $(f_{1,l_{1}},f_{2,l_{2}})$. We can naturally see that in order to increase the accuracy of the estimates of the harmonics present in the $i$-th dimension,  $\{f_{i,k}^{*}\}_{k=1}^{K}$, we must increase the length $L_{i}$ of the grid $\mathcal{F}_{i}$ used in building the subdictionary $\mathbf{D}_{i}$ so as to obtain a higher density grid.
\begin{figure}[htbp]
    \centering
    \includegraphics[width=0.5\linewidth, height = 0.55\linewidth]{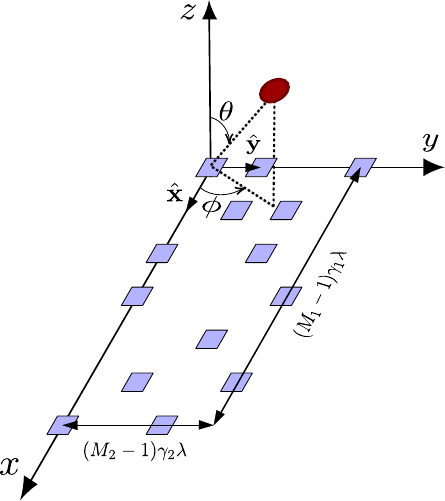} 
    \caption{A 2D $M$-element sparse array obtained by randomly subsampling elements from a URA with apertures in the $x$ and $y$ axes given by $(M_{1}-1)\gamma_{1}\lambda$ and $(M_{2}-1)\gamma_{2}\lambda$, respectively.}
    \label{fig:sparse_array}
\end{figure}
If we know randomly subsample $M$ elements from the the full URA while keeping the apertures in the $x$ and $y$ directions the same, as shown in Fig. \ref{fig:sparse_array}, then the signal model for the sparse array case can be formulated as
\begin{align}
	\mathbf{y}_{s} &= \mathbf{\Psi}\mathbf{y}
	 = \sum_{k=1}^{K}c_{k}\mathbf{\Psi} \mathbf{a}_{2}(f_{2,k}^{*}) \otimes \mathbf{a}_{1}(f_{1,k}^{*})+\mathbf{\Psi}\mathbf{n},
\end{align}
where $\mathbf{y}_{s} \in \mathbb{C}^{M}$ is the signal received on the $M$-element sparse array and $\mathbf{\Psi} \in \mathbb{R}^{M\times{M_{1}M_{2}}}$ is the subsampling matrix where $\mathbf{\Psi}(m, m_{1}+ m_{2}M_{1})=1$ if the $(m_{1}, m_{2})$-th element of the full URA is subsampled, and $\mathbf{\Psi}(m, m_{1}+ m_{2}M_{1})=0$ otherwise. The matrix $\mathbf{\Psi}$ is therefore a binary matrix with $M$ non-zero elements. The forward model for the URA in \eqref{eq:forward_model} is then updated to the sparse array case as follows
\begin{align}
	\mathbf{y}_{s}=\mathbf{\Psi}\mathbf{D}\mathbf{c}+\mathbf{\Psi}\mathbf{n}
	\triangleq \mathbf{D}_{s}\mathbf{c} + \mathbf{n}_{s},
	\label{eq:sparse_forward_model}
\end{align}
where we defined the row-subsampled dictionary $\mathbf{D}_{s} \triangleq \mathbf{\Psi}\mathbf{D} \in \mathbb{C}^{M\times L_{1}L_{2}}$ and $\mathbf{n}_{s} \triangleq \mathbf{\Psi}\mathbf{n}\in\mathbf{C}^{M}$. Finding an estimate of $\mathbf{c}$ in the forward model in \eqref{eq:sparse_forward_model} can be acomplished under the LASSO framework by solving the following optimization problem
\begin{align}
	\hat{\mathbf{c}}=\argmin_{\mathbf{c}} ||\mathbf{y}_{s}-\mathbf{D}_{s}\mathbf{c}||_{2}^{2}+\tau ||\mathbf{c}||_{1},
	\label{eq:lasso_objective}
\end{align}
where $\tau$ is a hyperparameter that controls the sparsity level of the solution $\hat{\mathbf{c}}$ and the first term in \eqref{eq:lasso_objective} is the data-fidelity term. Terminology-wise, it should be noted here that there are two distinct objects to whom the same qualifier ``sparse'' applies, namely the $M$-element sparse array itself that was obtained by subsampling from the URA, as well as the sparse solution $\hat{\mathbf{c}}$ to the problem in \eqref{eq:lasso_objective}. In the next section, we will review the computational complexity of three commonly used methods for solving the optimization problem in \eqref{eq:lasso_objective}, namely the ISTA, FISTA, and ADMM algorithms. In what follows, we will set $L= L_{1}L_{2}$.

\subsection{Sparse Recovery Methods: ISTA, FISTA and ADMM}
The ISTA solution to the problem in (\ref{eq:lasso_objective}) is obtained by directly applying the proximal-gradient descent method on the objective function and is given by the following fixed-point iterative scheme\cite{ista}
\begin{align}
	\mathbf{c}^{(t+1)}=S_{\kappa_{1}}\big( (\mathbf{I}_{L}-\mu \mathbf{D}_{s}^{H}\mathbf{D}_{s})\mathbf{c}^{(t)}+\mu\mathbf{D}_{s}^{H}\mathbf{y}_{s}\big),
	\label{eq:ista}
\end{align}
where $S_{\kappa}(z) = \exp(j\arg(z))\max(|z|-\lambda,0)$ is the soft-thresholding operator, $\mu = 1/\sigma_{max}(\mathbf{D}_{s})^{2}$ is the step-size, where $\sigma_{max}(\mathbf{D}_{s})$ is the largest singular value of $\mathbf{D}_{s}$, $\mathbf{I}_{L}$ is the size $L$ identity matrix, $\kappa_{1} = \mu \tau$ is the threshold level, and we assume a given initial value $\mathbf{c}^{(0)}$, which is typically set to the null vector. After running the iterative scheme in \eqref{eq:ista} for $T$ steps, the estimate $\hat{\mathbf{c}}$ to the problem in \eqref{eq:lasso_objective} is simply $\hat{\mathbf{c}} = \mathbf{c}^{(T)} $. As can be seen, the computational complexity of each iteration is dominated by the product $(\mathbf{I}_{L}-\mu \mathbf{D}_{s}^{H}\mathbf{D}_{s})\mathbf{c}^{(t)}$ and amounts to $\mathcal{O}(L^{2})$ operations. FISTA is an accelerated version of ISTA that employs an additional momentum term to increase the convergence rate of the ISTA algorithm, and is given by the following scheme \cite{fista}
\begin{align}
		&\mathbf{c}^{(t)}=S_{\kappa_{1}}\big( (\mathbf{I}_{L}-\mu \mathbf{D}_{s}^{H}\mathbf{D}_{s})\mathbf{z}^{(t)}+\mu\mathbf{D}_{s}^{H}\mathbf{y}_{s}\big) \label{eq:fista}\\
		&\alpha^{(t+1)} = \big(\sqrt{1+4(\alpha^{(t)})^2} + 1 \big)/2 \\
		&\mathbf{z}^{(t+1)} = \mathbf{c}^{(t)} + \dfrac{\alpha^{(t)}-1}{\alpha^{(t+1)}}(\mathbf{c}^{(t)}-\mathbf{c}^{(t-1)}),
\end{align}
along with an initialization for $\mathbf{c}^{(0)}$, where $\mathbf{z}^{(1)}$ is typically set to the same value as $\mathbf{c}^{(0)}$, and $\alpha^{(0)}=1$. Similarly to ISTA, the dominant term in the FISTA iterative scheme is the matrix product $(\mathbf{I}_{L}-\mu \mathbf{D}_{s}^{H}\mathbf{D}_{s})\mathbf{z}^{(t)}$ in \eqref{eq:fista}  which also amounts to $\mathcal{O}(L^{2})$ per iteration. Finally, another popular algorithm that exhbits a competitve performance \cite{num_iterations} against ISTA and FISTA for solving the optimization problem in \eqref{eq:lasso_objective} is ADMM, which is obtained by alternating between gradient descent on the primal variables corresponding to the objective function in \eqref{eq:lasso_objective} and gradient ascent on the dual variable, resulting in the following three-step method \cite{admm}
\begin{align}
	&\mathbf{c}^{(t+1)}=(\mathbf{D}_{s}^{H}\mathbf{D}_{s}+\rho \mathbf{I}_{L})^{-1}(\mathbf{D}_{s}^{H}\mathbf{y}_{s}+\rho(\mathbf{z}^{(t)}-\mathbf{v}^{(t)})) \label{eq:admm}\\
	&\mathbf{z}^{(t+1)}=S_{\kappa_{2}}(\mathbf{c}^{(t+1)}+\mathbf{v}^{(t)})\\
	&\mathbf{v}^{(t+1)}=\mathbf{v}^{(t)}+\mathbf{c}^{(t+1)}-\mathbf{z}^{(t+1)},
\end{align}
where $\rho>0$ is the Lagrange hyperparameter that influences the convergence rate of the algorithm, $\kappa_{2}=\rho\tau$ is the threshold level, and the initial values are specified by $\mathbf{z}^{(0)}$ and $\mathbf{v}^{(0)}$, typically both set to the null vector. Since the term $(\mathbf{D}_{s}^{H}\mathbf{D}_{s}+\rho \mathbf{I}_{L})^{-1}\mathbf{D}_{s}^{H}\mathbf{y}_{s}$ in \eqref{eq:admm} can be precomputed and stored, we can infer that the dominant computational term in this scheme is the matrix product $(\mathbf{D}_{s}^{H}\mathbf{D}_{s}+\rho \mathbf{I}_{L})^{-1}(\mathbf{z}^{(t)}-\mathbf{v}^{(t)})$ which, similarly to ISTA and FISTA, also amounts to $\mathcal{O}(L^{2})$ operations per iteration. 
\section{Block Circulant Matrices with Circulant Blocks}
\label{using bccbs}
\subsection{Block-Circulant Matrices with Circulant Blocks}
As was shown in \cite{tlista}, if the grid $\mathcal{F}_{i}$ used to build the subdictionary $\mathbf{D}_{i}$ is uniform,  i.e. 
\begin{align}
f_{i,l_{i}}=-\gamma_{i}+ l_{i}(2\gamma_{i}/L_{i}),
\label{eq:uniform_grid}
\end{align} 
for $l_{i}=0, 1, \hdots, L_{i}-1$, then the Gram of the subdictionary $\mathbf{D}_{i}$, i.e. $\mathbf{D}_{i}^{H}\mathbf{D}_{i}$, will be a Toeplitz matrix. Moreover, if, in addition to the uniformity constraint on the harmonics grid, we also have $\gamma_{i}=1/2$, then it can be shown \cite{circadmm} by substituting \eqref{eq:uniform_grid} into \eqref{eq:subdictionary} that the Gram of the subdictionary $\mathbf{D}_{i}$ will reduce to a circulant matrix. If we denote the $m_{i}$-th row of $\mathbf{D}_{i}$ as $\mathbf{d}_{i,m_{i}}$, we can similarly show by substitution that $\mathbf{d}_{i,m_{i}}^{H}\mathbf{d}_{i,m_{i}}$ is also a circulant matrix. The condition on $\gamma_{1}=1/2$ or $\gamma_{2}=1/2$ amounts to having the $x$ or $y$ coordinates, respectively, of the array elements specified in integers units of $\lambda/2$, which is typically the case in practice.  We will henceforth set $\gamma_{1} = \gamma_{2} = 1/2$ along with the assumption of a uniform grid for $\mathcal{F}_{1}$ and $\mathcal{F}_{2}$. With these conditions, we can now show that $\mathbf{D}_{s}^{H}\mathbf{D}_{s}$ is a BCCB matrix. We define the indicator function $\mathds{1}_{s}(m_{1},m_{2})$ that is equal to $1$ if the the sparse array contains an element with coordinates $(m_{1}\lambda/2, m_{2}\lambda/2,0)$, and $0$ otherwise, we then have
\begin{align}
	&\mathbf{D}_{s}^{H}\mathbf{D}_{s}=(\mathbf{\Psi}\mathbf{D})^{H}\mathbf{\Psi}\mathbf{D} =(\mathbf{\Psi}(\mathbf{D}_{2}\otimes \mathbf{D}_{1}))^{H} (\mathbf{\Psi}(\mathbf{D}_{2}\otimes \mathbf{D}_{1})) \nonumber \\
	&=\sum_{m_{1}=0}^{M_{1}-1}\sum_{m_{2}=0}^{M_{2}-1} \mathds{1}_{s}(m_{1},m_{2}) (\mathbf{d}_{2,m_{2}}^{H} \otimes  \mathbf{d}_{1,m_{1}}^{H}) (\mathbf{d}_{2,m_{2}} \otimes  \mathbf{d}_{1,m_{1}}) \nonumber \\
	&=\sum_{m_{1}=0}^{M_{1}-1}\sum_{m_{2}=0}^{M_{2}-1} \mathds{1}_{s}(m_{1},m_{2}) (\mathbf{d}_{2,m_{2}} \mathbf{d}_{2,m_{2}}^{H}) \otimes  (\mathbf{d}_{1,m_{1}} \mathbf{d}_{1,m_{1}}^{H}),\label{eq:bccb_sum}
\end{align}
and since $\mathbf{d}_{i,m_{i}} \mathbf{d}_{i,m_{i}}^{H}$ is a circulant matrix for $i=1,2$, $(\mathbf{d}_{1,m_{1}} \mathbf{d}_{1,m_{1}}^{H}) \otimes  (\mathbf{d}_{2,m_{2}} \mathbf{d}_{2,m_{2}}^{H})$  will be a BCCB. As BCCBs are closed under addtion, i.e. the sum of two BCCBs is also a BCCB, we can then conclude that the sum in \eqref{eq:bccb_sum}, and hence $\mathbf{D}_{s}^{H}\mathbf{D}_{s}$ will be a BCCB. The following properties \cite{trapp} of BCCB matrices are listed for subsequent developments
\begin{enumerate}
\renewcommand{\labelenumi}{(\roman{enumi}).}
\item If $\mathbf{R}_{1},\mathbf{R}_{2}$ are BCCBs, then so is $\alpha_{1}\mathbf{R}_{1}+\alpha_{2}\mathbf{R}_{1}$, where $\alpha_{1}, \alpha_{2} \in \mathbb{C}$.
\item BCCBs are closed under inversion.
\item  The identity $\mathbf{I}_{L}$ matrix of size $L=L_{1}L_{2}$ is a BCCB with $L_{2}$ blocks of shape $L_{1}$ by $L_{1}$.
\item If $\mathbf{R}$ is a BCCB with $L_{2}$ blocks of shape $L_{1}$ by $L_{1}$, then $\mathbf{R}$ can be fully charachterized by a sequence $\mathcal{S}_{\mathbf{R}} =\{\mathbf{r}_{l_{2}}\}_{l_{2}=0}^{L_{2}}$ of $L_{2}$ vectors of size $L_{1}$, i.e. $\mathbf{r}_{l_{2}} \in \mathbf{C}^{L_{1}}$. Moreover, denoting the first column of $\mathbf{R}$ by $\mathbf{r}$, then we have $\mathbf{r}_{l_{2}}(l_{1}) = \mathbf{r}(l_{1}+l_{2}L_{1})$.
\end{enumerate}
Using the above mentioned properties, it can be easily shown that both  $(\mathbf{I}_{L}-\mu \mathbf{D}_{s}^{H}\mathbf{D}_{s})$ and $(\mathbf{D}_{s}^{H}\mathbf{D}_{s}+\rho \mathbf{I}_{L})^{-1}$ are also BCCB matrices. Additionally, one of the most important propreties of BCCB matrices is that they are diagonalizable using discrete 
\begin{figure*}[htbp]
    \centering
    \subcaptionbox{ISTA\label{fig:plot1}}{%
        \includegraphics[width=0.3\textwidth]{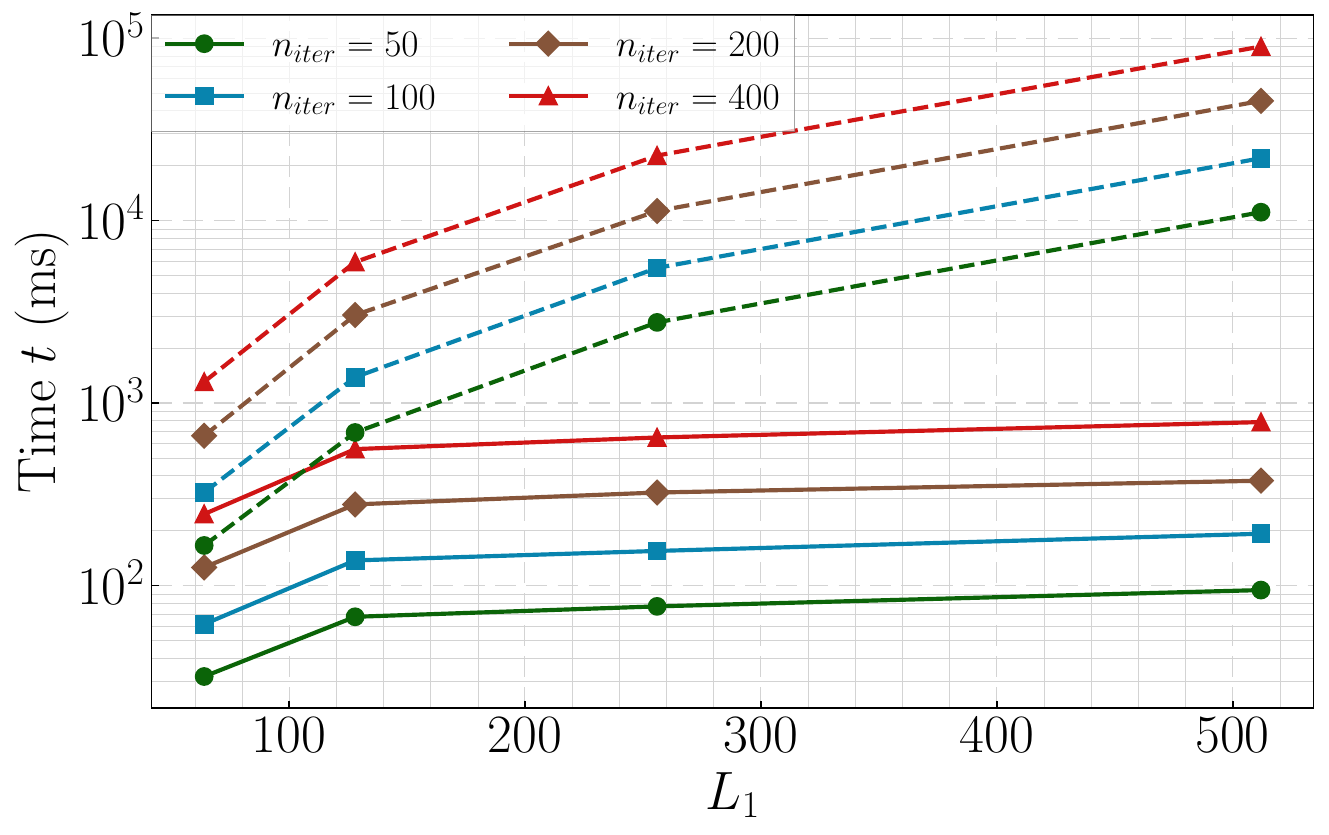}}
    \hfill
    \subcaptionbox{FISTA\label{fig:plot2}}{%
        \includegraphics[width=0.3\textwidth]{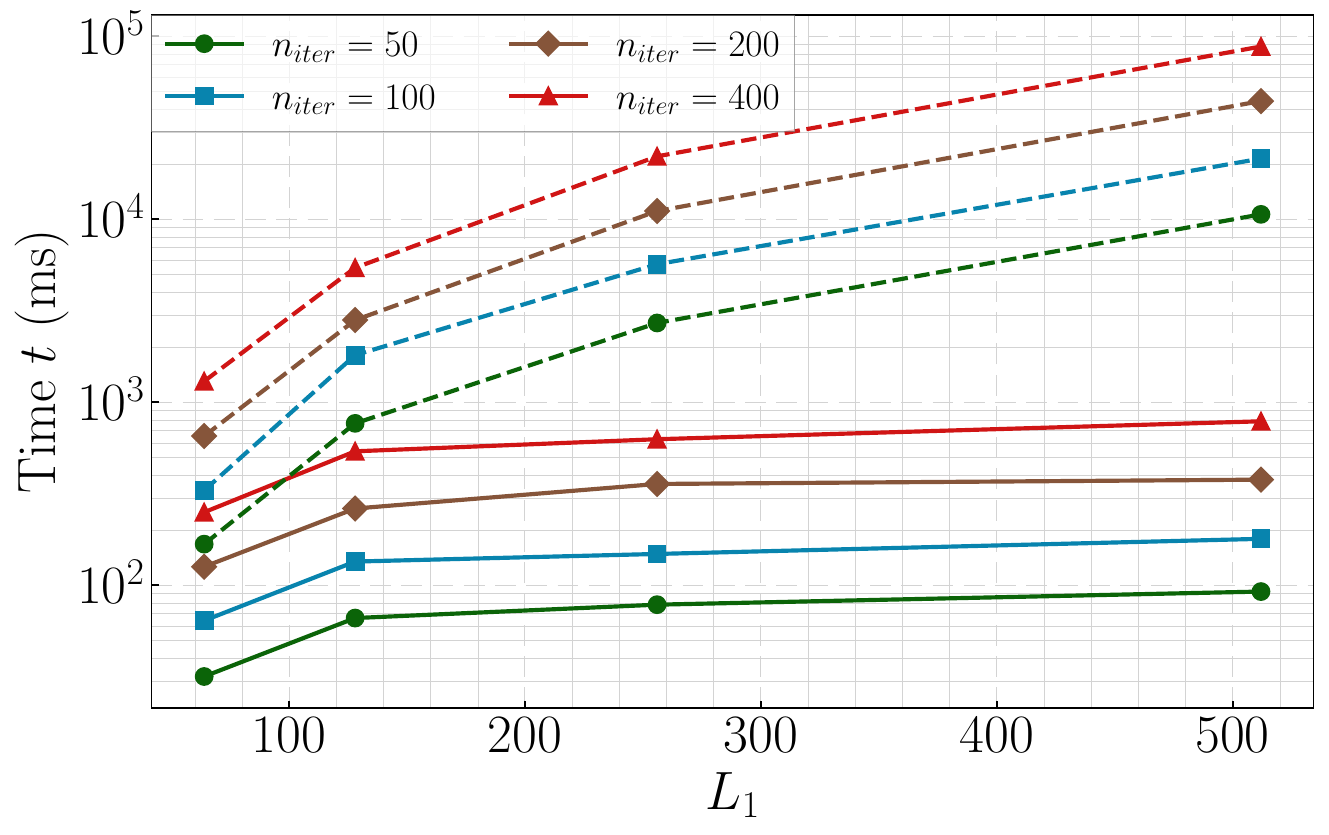}}
    \hfill
    \subcaptionbox{ADMM \label{fig:plot3}}{%
        \includegraphics[width=0.3\textwidth]{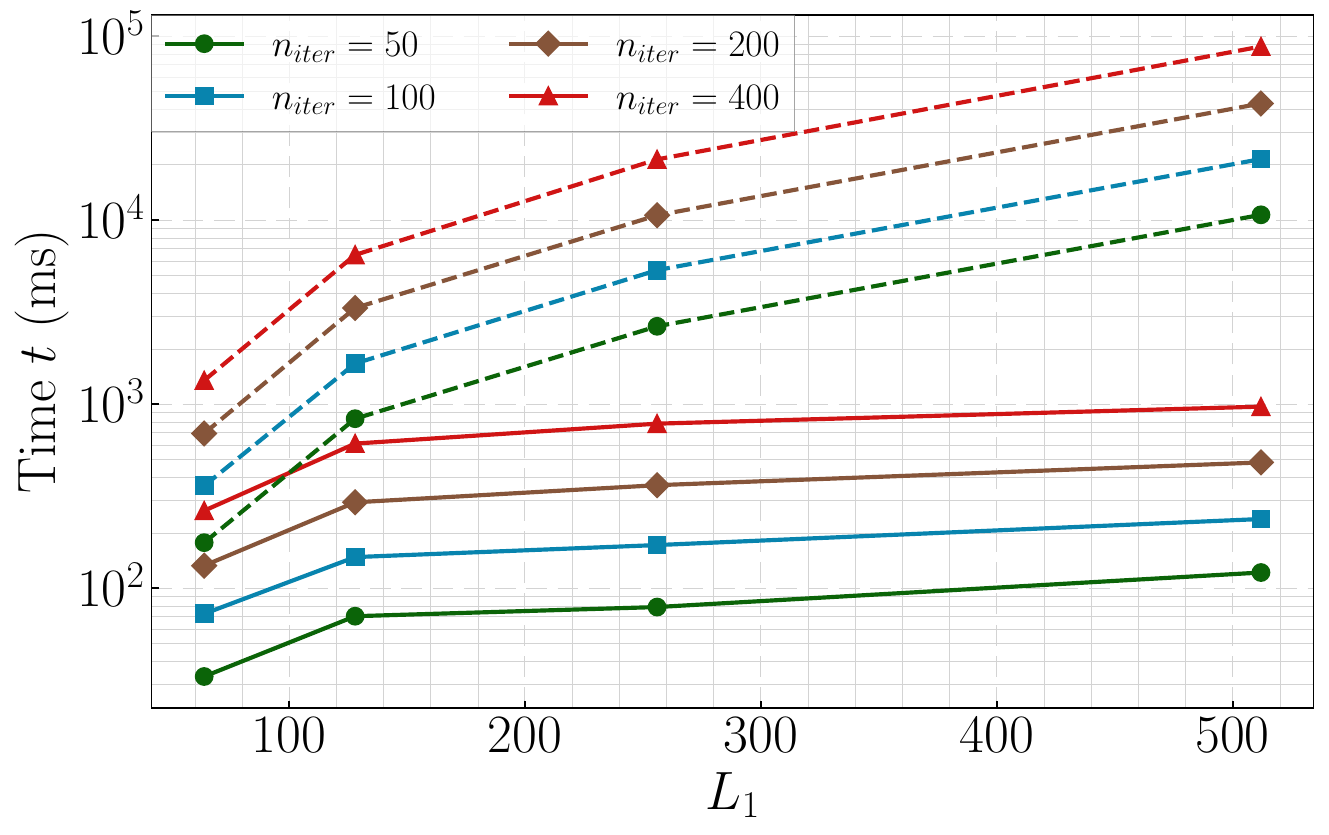}}
    
    \caption{Comparison of the runtime in milliseconds (ms) for the regular implementation $t_{reg}$ (in dashed lines) against the runtime for the fast implementation $t_{fast}$ (in continuous lines) for ISTA, FISTA and ADMM.}
    \label{fig:overall}
\end{figure*}
\begin{table*}[h!]
\centering
\begin{subtable}[t]{0.30\textwidth}
\centering
\begin{tabular}{|c|c|c|c|c|c|}
\cline{3-6}
\multicolumn{2}{c|}{} & \multicolumn{4}{c|}{$L_{1}$} \\
\cline{3-6}
\multicolumn{1}{c}{} & \multicolumn{1}{c|}{} & 64 & 128 & 256 & 512 \\
\hline
\multirow{4}{*}{\rotatebox{90}{$n_{iter}$}}& 50 & 2.67 & 2.18 & 2.31 & 3.13 \\
\cline{2-6}
& 100 & 4.97 & 4.59 & 4.77 & 10.61 \\
\cline{2-6}
& 200 & 4.57 & 4.87 & 10.62 & 13.70 \\
\cline{2-6}
& 400 & 7.40 & 5.43 & 25.40 & 62.82 \\
\hline
\end{tabular}
\caption{$\epsilon_{r}\times 10^{-14}$}
\end{subtable}
\hfill
\begin{subtable}[t]{0.30\textwidth}
\centering
\begin{tabular}{|c|c|c|c|c|c|}
\cline{3-6}
\multicolumn{2}{c|}{} & \multicolumn{4}{c|}{$L_{1}$} \\
\cline{3-6}
\multicolumn{1}{c}{} & \multicolumn{1}{c|}{} & 64 & 128 & 256 & 512 \\
\hline
\multirow{4}{*}{\rotatebox{90}{$n_{iter}$}} & 50 & 2.35 & 1.45 & 0.63 & 0.61 \\
\cline{2-6}
& 100 & 2.90 & 1.88 & 1.02 & 0.66 \\
\cline{2-6}
& 200 & 4.20 & 1.33 & 1.48 & 1.67 \\
\cline{2-6}
& 400 & 3.47 & 2.17 & 1.93 & 2.69 \\
\hline
\end{tabular}
\caption{$\epsilon_{r}\times 10^{-14}$}
\end{subtable}
\hfill
\begin{subtable}[t]{0.30\textwidth}
\centering
\begin{tabular}{|c|c|c|c|c|c|}
\cline{3-6}
\multicolumn{2}{c|}{} & \multicolumn{4}{c|}{$L_{1}$} \\
\cline{3-6}
\multicolumn{1}{c}{} & \multicolumn{1}{c|}{} & 64 & 128 & 256 & 512 \\
\hline
\multirow{4}{*}{\rotatebox{90}{$n_{iter}$}} & 50 & 1.21 & 2.77 & 6.03 & 7.77 \\
\cline{2-6}
& 100 & 1.10 & 1.94 & 5.52 & 12.00 \\
\cline{2-6}
& 200 & 1.69 & 1.80 & 7.35 & 16.87 \\
\cline{2-6}
& 400 & 1.56 & 2.28 & 7.66 & 23.46 \\
\hline
\end{tabular}
\caption{$\epsilon_{r}\times 10^{-10}$}
\end{subtable}
\caption{The relative reconstruction error $\epsilon_{r}$ between the regular and fast implementation for (a) ISTA, (b) FISTA, and (c) ADMM.}
\label{table}
\end{table*}
Fourier tranform (DFT) matrices. Denoting the $L_{i}$ point DFT matrix by $\mathbf{F}_{L_{i}}$, it can be shown  \cite{vescovo} that if $\mathbf{R}$ is BCCB with $L_{2}$ blocks of shape $L_{1}$ by $L_{1}$, then it is diagonalizable under the following form
\begin{align}
\mathbf{R}=(\mathbf{F}_{L_{2}}^{-1} \otimes \mathbf{I}_{L_{1}})\mathbf{\Gamma}^{-1}\mathbf{\Lambda}\mathbf{\Gamma}(\mathbf{F}_{L_{2}} \otimes \mathbf{I}_{L_{1}}), \label{eq:diagonalization}
\end{align}
where $\mathbf{\Gamma}\in \mathbb{C}^{L\times L}$ is a block-diagonal matrix with $L_{2}$ blocks and is explicitly given by
\begin{align}
\mathbf{\Gamma} = \textrm{diag}(\underbrace{\mathbf{F}_{{L}_{1}}, \mathbf{F}_{L_{1}}, \hdots, \mathbf{F}_{L_{1}}}_{L_{2}\text{ blocks}}),  
\end{align}
along with its inverse $\mathbf{\Gamma}_{L_{1}}^{-1} = \textrm{diag}(\mathbf{F}_{{L}_{1}}^{-1}, \mathbf{F}_{L_{1}}^{-1}, \hdots, \mathbf{F}_{L_{1}}^{-1})$, and $\mathbf{\Lambda} \in \mathbb{C}^{L\times L}$ is a block-diagonal matrix where the blocks are themselves diagonal matrices, and is given by
\begin{align}
\mathbf{\Lambda} = \textrm{diag}(\underbrace{\textrm{diag}(\boldsymbol{\lambda}_{0}), \textrm{diag}(\boldsymbol{\lambda}_{1}), \hdots, \textrm{diag}(\boldsymbol{\lambda}_{L_{2}-1})}_{L_{2}\text{ blocks}}),
\end{align}
where 
\begin{align}
	\boldsymbol{\lambda}_{l_{2}'}=\mathbf{F}_{L_{1}}\bigg( \sum_{l_{2}=0}^{L_{2}-1}\mathbf{r}_{l_{2}}\exp(-j2\pi \frac{l_{2}}{L_{2}}l_{2}') \bigg), \label{eq:lambda}
\end{align}
for $l_{2}' = 0, 1, \hdots, L_{2}-1$, and where $\mathbf{r}_{l_{2}}$ is defined in (iv) among the previously listed properties of BCCB matrices.  The quantity inside the parentheses in \eqref{eq:lambda} can be interpereted as the $L_{2}$-point DFT of the vector sequence $\{\mathbf{r}_{l_{2}}\}_{l_{2}=0}^{L_{2}}$ that fully charachterizes $\mathbf{R}$. 
\subsection{Efficient Computation of BCCB-Vector Products}
Given a vector $\mathbf{c}\in \mathbb{C}^{L}$ partitioned into $L_{2}$ contiguous blocks $\mathbf{c}_{l_{2}}\in \mathbb{C}^{L_{1}}$, with $l_{2}=0,\hdots, L_{2}-1$, we can write
\begin{align}
	\mathbf{R}\mathbf{c} &= (\mathbf{F}_{L_{2}}^{-1} \otimes \mathbf{I}_{L_{1}})\mathbf{\Gamma}^{-1}\mathbf{\Lambda}\mathbf{\Gamma}(\mathbf{F}_{L_{2}} \otimes \mathbf{I}_{L_{1}}) \textrm{vec}(\mathbf{C}),
\end{align}
where we defined $\mathbf{C} \in \mathbb{C}^{L_{1}\times L_{2}}$ as $\mathbf{C} = [\mathbf{c}_{0}, \mathbf{c}_{1}, \hdots, \mathbf{c}_{L_{2}-1}]$. By exploiting both the block-diagonal nature of $\mathbf{\Gamma}$ and $\mathbf{\Gamma}^{-1}$, the purely diagonal nature of $\mathbf{\Lambda}$, along with the relationship $\mathbf{A}\otimes \mathbf{B}\vect(\mathbf{C})=\vect(\mathbf{B}\mathbf{C}\mathbf{A}^{T})$,  we can write
\begin{align}
	\mathbf{R}\mathbf{c} &= (\mathbf{F}_{L_{2}}^{-1} \otimes \mathbf{I}_{L_{1}})\mathbf{\Gamma}^{-1}\mathbf{\Lambda}\mathbf{\Gamma}\textrm{vec}(\mathbf{I}_{L_{1}}\mathbf{C} \mathbf{F}_{L_{2}}^{T} ) \nonumber\\
	&= (\mathbf{F}_{L_{2}}^{-1} \otimes \mathbf{I}_{L_{1}})\mathbf{\Gamma}^{-1}\mathbf{\Lambda}\textrm{vec}(\mathbf{F}_{L_{1}}\mathbf{C} \mathbf{F}_{L_{2}}^{T} ) \nonumber\\
	&= (\mathbf{F}_{L_{2}}^{-1} \otimes \mathbf{I}_{L_{1}})\mathbf{\Gamma}^{-1}_{L_{1}}\textrm{vec}(\mathbf{\Omega}\odot\mathbf{F}_{L_{1}}\mathbf{C} \mathbf{F}_{L_{2}}^{T} ) \nonumber\\
	&= (\mathbf{F}_{L_{2}}^{-1} \otimes \mathbf{I}_{L_{1}})\textrm{vec}(\mathbf{F}_{L_{1}}^{-1}(\mathbf{\Omega}\odot\mathbf{F}_{L_{1}}\mathbf{C} \mathbf{F}_{L_{2}}^{T}) ) \nonumber\\
	&= \textrm{vec}\big(\mathbf{F}_{L_{1}}^{-1}(\mathbf{\Omega}\odot\mathbf{F}_{L_{1}}\mathbf{C} \mathbf{F}_{L_{2}}^{T})(\mathbf{F}_{L_{2}}^{-1})^{T}\big), \label{eq:Rc}
\end{align}
where $\mathbf{\Omega}\in \mathbb{C}^{L_{1}\times L_{2}}$ is defined as $\mathbf{\Omega}=[\boldsymbol{\lambda}_{0}, \boldsymbol{\lambda}_{1}, \hdots, \boldsymbol{\lambda}_{L_{2}-1}]$ and $\odot$ is the Hadamard product. Since the quantity $\mathcal{F}\{\mathbf{C}\}=\mathbf{F}_{L_{1}}\mathbf{C} \mathbf{F}_{L_{2}}^{T}$ that appears in \eqref{eq:Rc} corresponds to the 2D DFT of $\mathbf{C}$, along with $\mathbf{F}_{L_{1}}^{-1}(\mathbf{\Omega}\odot\mathcal{F}\{\mathbf{C}\})(\mathbf{F}_{L_{2}}^{-1})^{T}$ corresponding to the 2D IDFT of $\mathbf{\Omega}\odot\mathcal{F}\{\mathbf{C}\}$, we can conclude that a product of the form $\mathbf{R}\mathbf{c}$ where $\mathbf{R}$ is a BCCB, such as the product present in \eqref{eq:ista}, \eqref{eq:fista} or \eqref{eq:admm}, can be efficiently computed through the use of 2D FFTs in $\mathcal{O}(L\log(L))$ instead of $ \mathcal{O}(L^{2})$ operations. 
\subsection{Experimental Results}
We set $M_{1}=51$ and $M_{2}=16$ and randomly sample 40 elements from the URA while preserving its aperture of $50 \lambda/2$ by $15 \lambda/2$. Next, we vary $L=L_{1}L_{2}$ by fixing $L_{2}=32$ and varying $L_{1} \in \{64, 128, 256, 512\}$. For each value of $L$, we compare the runtime $t_{reg}$ of the regular implementation of ISTA, FISTA, and ADMM against the runtime $t_{fast}$ of the fast implementation  based on 2D FFTs. For both implementations, we vary the number of iterations $n_{iter} \in \{50, 100, 200, 400\}$. The noise variance $\sigma^{2}$ in \eqref{eq:signal_model} is set so as to obtain a signal-to-noise ratio (SNR) of $15$ dB, the number of sources $K$ is randomly generated between 1 and 10, and each harmonic tuple $(f_{1,k}^{*}, f_{2,k}^{*})$ is randomly generated from $[-0.5, 0.5[ \times [-0.5, 0.5[ $. For each combination of $(L, n_{iter})$, we report the average value over 10 trials of $t_{reg}$, $t_{fast}$ and the relative reconstruction error $\epsilon_{r}=||\hat{\mathbf{c}}_{reg}-\hat{\mathbf{c}}_{fast}||_{2}/||\hat{\mathbf{c}}_{reg}||_{2}$. The experiments are conducted using an AMD EPYC 7343 @ 3.20 GHz CPU with 32 GB of RAM and the 2.5.1 PyTorch library. The runtime results are shown in Fig. \ref{fig:overall} whereas the reconstruction error results are shown in Table \ref{table}. 

\section{Conclusion}
\label{conclusion}
We presented an efficient implementation of the ISTA, FISTA and the ADMM algorithms for sparse recovery of harmonics using 2D sparse arrays. By exploting the BCCB matrix structure that arise in these iterative methods, we showed that the computational complexity of the matrix-vector products can be reduced from $\mathcal{O}(L^{2})$ to $\mathcal{O}(L\log(L))$ operations per iteration by using 2D FFTs.

\bibliographystyle{IEEEbib}
\bibliography{refs}

\end{document}